%% file: main.tex
\documentclass[conference]{IEEEtran}
\IEEEoverridecommandlockouts
\usepackage{cite}
\usepackage{amsmath,amssymb,amsfonts,amsthm}
\usepackage{mathtools}
\usepackage{makecell}
\usepackage{graphicx}
\usepackage{textcomp}
\usepackage{xcolor}
\usepackage{booktabs}
\usepackage{url}
\usepackage{multirow}
\usepackage{algorithm}
\usepackage{algpseudocode}
\usepackage{bm}
\usepackage{placeins}
\usepackage{float}
\usepackage{mathptmx}
\usepackage{times}
\DeclareMathAlphabet{\mathcal}{OMS}{cmsy}{m}{n}
\graphicspath{{figs/}}

\usepackage[colorlinks=true,linkcolor=blue,citecolor=blue,urlcolor=blue]{hyperref}
\hypersetup{pdftitle={Model-Free UPS Control Tuning for Improved Low-Voltage Ride-Through in AI Data Centers},pdfauthor={Haowen Xu and Xin Chen}}
\newcommand{\Wrap}[2][0]{\parbox[t]{\dimexpr\linewidth-#1\dimexpr\algorithmicindent\relax\relax}{#2\strut}}
\allowdisplaybreaks

\begin{document}
\bstctlcite{IEEEexample:BSTcontrol}

\title{Model-Free UPS Control Tuning for Improved Low-Voltage Ride-Through in AI Data Centers}


\author{\IEEEauthorblockN{Haowen Xu}
\IEEEauthorblockA{\textit{Department of Electrical and Computer Engineering} \\
\textit{Texas A\&M University}\\
College Station, TX, USA \\
haowen\_xu@tamu.edu}
\and
\IEEEauthorblockN{Xin Chen}
\IEEEauthorblockA{\textit{Department of Electrical and Computer Engineering} \\
\textit{Texas A\&M University}\\
College Station, TX, USA \\
xin\_chen@tamu.edu}

\thanks{This work was supported in part by NSF CAREER 2541998, in part by the Power Systems Engineering Research Center (PSERC) Project S-118, and in part by the Consortium on AI and Large Flexible Load (CALL) at Texas A\&M University. }
}

\maketitle

\input{Abstract}

\input{Introduction}

\input{Problem_Setup}

\input{Problem_Formulation}

\input{Algorithm_Design}

\input{Case_Study_and_Results}

\input{Conclusion}

\FloatBarrier
\bibliographystyle{IEEEtran}
\bibliography{refs}

\end{document}

%% file: Abstract.tex
\begin{abstract}

Voltage sags can trigger protection trips in uninterruptible power supply (UPS) systems, abruptly reducing the power that AI data centers draw from the grid. This paper improves the low-voltage ride-through (LVRT) capability of AI data centers equipped with double-conversion UPS systems by tuning their existing converter controls. We formulate a bound-constrained optimization problem to maximize the ride-through area, defined as the integral of ride-through time over a range of retained grid voltages. To solve this problem, we develop a batched projected zeroth-order optimization algorithm with adaptive moment estimation (BPZO-Adam). 
The algorithm evaluates candidate control settings using only simulated ride-through times, making it suitable for vendor-supplied black-box models. Simulations using the TAMU-ERCOT electromagnetic transient (EMT) AI data center model show that tuning converter control parameters increases the ride-through area by 10.6\%. In a separate study, tuning the controls yields a larger ride-through area than increasing the DC-link capacitance tenfold.

\end{abstract}



%% file: Introduction.tex
\begin{figure*}[t]
\centering
\includegraphics[width=0.9\textwidth]{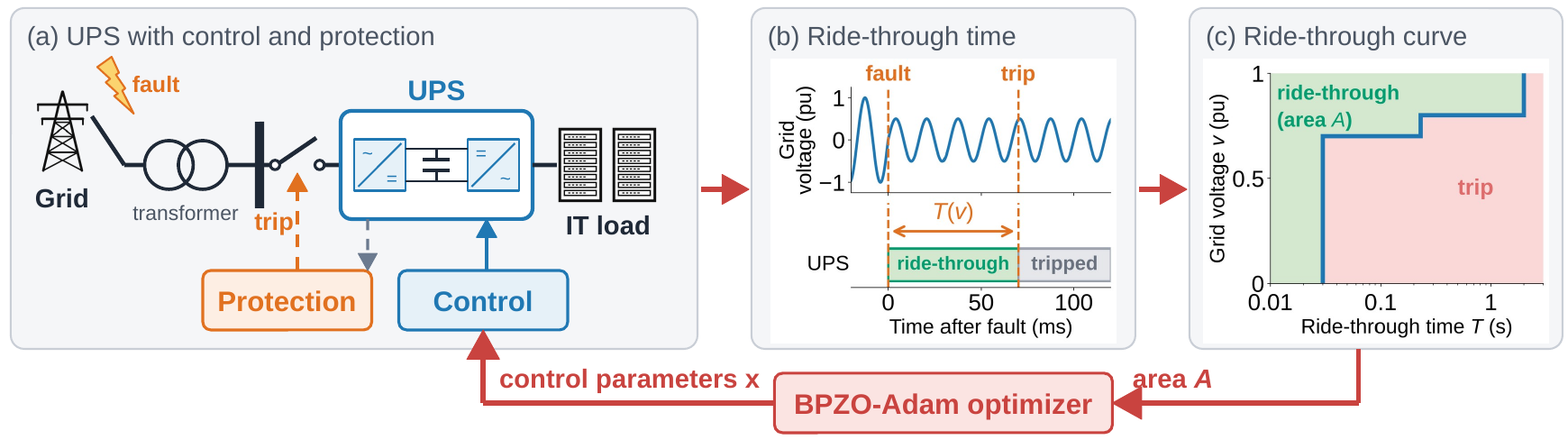}
\caption{The proposed model-free, simulation-based UPS control optimization framework for enhancing AI data center LVRT. 
(a) Double-conversion UPS-based AI data center with converter controls and protection. (b) Ride-through time $T(v)$ under a grid voltage sag $v$. (c) Ride-through curve over all grid voltage sag levels and its ride-through area $A$. The BPZO-Adam optimizer optimizes the control parameters $\mathbf{x}$ to maximize the ride-through area $A$.}
\label{fig:frame}
\end{figure*}

\section{Introduction}\label{sec:intro}

AI data centers are among the fastest-growing loads on the grid.
They are predominantly interfaced with the grid through power electronics, which are sensitive to grid-side voltage and frequency disturbances \cite{chen2026nexus}. AI servers receive power through uninterruptible power supply (UPS) systems. During a severe grid voltage dip, UPS protection can switch the unit to battery operation mode within milliseconds
while ceasing to draw power from the grid.
As a result, a grid fault can trigger this response at multiple data centers, causing a sudden loss of substantial loads seen by the grid. In July 2024, a single 230-kV line fault led to the loss of approximately 1.5 GW of data center load \cite{nerc_incident}, and another event in July 2026 resulted in a large load loss of more than 3 GW \cite{nerc_insight2026}. Abrupt load losses of this magnitude greatly threaten power system stability and reliability \cite{nerc_incident,chertkov2026gw}.



To prevent such losses, large data center loads are expected to remain connected to the grid during voltage dips, an ability known as low-voltage ride-through (LVRT) \cite{nerc_alert}. Specifically, LVRT requirements define a voltage-time curve that specifies the minimum ride-through duration at each retained voltage level \cite{nogrr282}. Ride-through requirements have been established for inverter-based resources  \cite{prc029}, while recent  initiatives by the North American Electric Reliability Corporation (NERC) and the Electric Reliability Council of Texas (ERCOT) address ride-through for large computational loads \cite{nerc_alert,nogrr282}. A key challenge in data center LVRT is to keep the UPS units connected to the grid throughout voltage sags
while maintaining a reliable power supply to the servers.



Existing efforts for enhancing AI data center LVRT range from adding hardware support to  facility-level control design. Hardware-based approaches include battery energy storage systems (BESSs), static synchronous compensators with supercapacitors (E-STATCOMs), and reconfigured DC distribution architectures~\cite{elliott2026}.
Substation braking resistors have also been proposed to absorb surplus power following an abrupt loss of load \cite{ghosh2025}. These measures require additional equipment or changes to the power delivery architecture. In terms of control-based approaches, Xie \emph{et al.} \cite{xie2026enhancing} developed centralized and decentralized controllers to coordinate on-site flexible energy resources for data center voltage regulation. Wang \emph{et al.} \cite{wang2026solvrt} used formal methods to synthesize an LVRT controller. Both studies represent data center resources as power injections in a linearized distribution network and use simplified disconnection rules. As a result, their models neglect the converter control dynamics and protection logic that govern whether and when the UPS rectifier trips during a voltage sag.


The LVRT capability of a double-conversion UPS depends on how its converter controls interact with its protection logic. The rectifier and inverter operate under cascaded control loops (Fig.~\ref{fig:frame}(a)). During a voltage sag, the rectifier increases its input current, subject to its current limit, to support the DC-link voltage. The converter control parameters shape the current response and the depth and duration of the DC-link voltage drop, which in turn influence whether and when protection operates (Fig.~\ref{fig:frame}(b)). 
This interaction shapes the UPS ride-through curve, which identifies the combinations of voltage sag depth and duration that the UPS can withstand (Fig.~\ref{fig:frame}(c)). Field observations also illustrate how UPS protection and control settings can affect the magnitude of load loss during voltage disturbances~\cite{nerc_insight2026}.
 Although tuning UPS converter controls is critical to LVRT capability, its effect on the UPS ride-through curve remains unexplored in existing studies.



Tuning UPS converter controls to improve LVRT poses two challenges. First, to protect proprietary design information, UPS manufacturers typically provide black-box electromagnetic transient (EMT) models that limit access to the governing equations needed for model-based tuning~\cite{wang2026tpwrs}. 
Second, small changes in control parameters can alter whether or when protection trips, causing discontinuous changes in the simulated ride-through time. 
These challenges motivate the use of zeroth-order optimization (ZO) \cite{liu2020}, a model-free approach that uses perturbed objective function evaluations for optimization without requiring analytical gradients~\cite{chen2021,xu_zoadam}.

To address these challenges, this paper develops a batched projected ZO with adaptive moment estimation (BPZO-Adam) algorithm that evaluates candidate UPS converter control parameters through EMT simulations and optimizes them to improve LVRT in AI data centers. This model-free, simulation-based LVRT optimization framework is illustrated in Fig. \ref{fig:frame}.
The main contributions of this paper are summarized below:
\begin{itemize}
    \item [(1)] \textbf{LVRT enhancement via UPS control tuning.} To our knowledge, this is the first study to optimize UPS converter control parameters specifically to improve LVRT in AI data centers using existing UPS hardware.
    \item [(2)]  \textbf{Black-box optimization with EMT simulations.} We develop BPZO-Adam to tune UPS converter controls using only the ride-through times simulated by vendor-supplied EMT models. The method requires neither access to the detailed converter models nor analytical gradients.
    \item [(3)] \textbf{Practical guidance on controller tuning.} From high-fidelity EMT case studies, optimized UPS converter controls yielded longer ride-through times than increasing the DC-link capacitance tenfold, underscoring the potential of control tuning to improve LVRT performance.
\end{itemize}




%% file: Problem_Setup.tex
\section{System Description}\label{sec:setup}

\subsection{Uninterruptible Power Supply Structure and Model}\label{sec:ups}
\begin{figure}[t]
\centering
\includegraphics[width=\columnwidth]{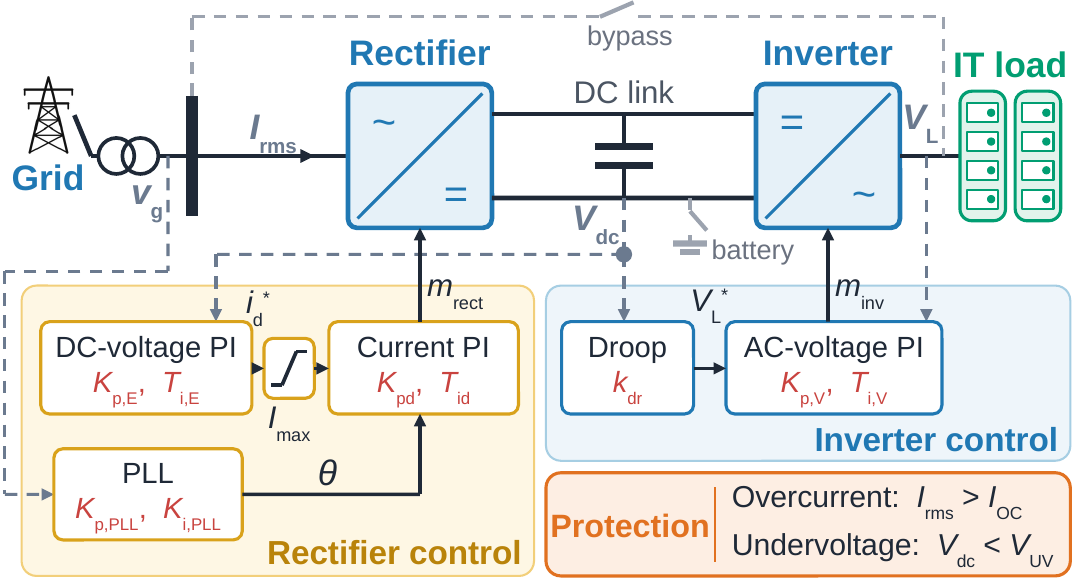}
\caption{Structure, control loops, and protection of the double-conversion UPS. The bypass and battery paths are disabled during the ride-through period.}
\label{fig:ups}
\end{figure}

Fig.~\ref{fig:ups} shows the structure of a double-conversion UPS connected to the grid through a step-down transformer \cite{tamu_model}. The rectifier converts the AC input into a regulated DC-link voltage, and the inverter supplies the IT load with a regulated AC voltage. The bypass path and battery remain open, so LVRT capability reflects only the converters and their control.

The rectifier adopts the standard cascaded control of a grid-connected converter \cite{yazdani2010}. A phase-locked loop (PLL) with proportional and integral gains $K_{p,\mathrm{PLL}}$ and $K_{i,\mathrm{PLL}}$ tracks the grid phase angle. An outer DC-voltage proportional-integral (PI) loop with proportional gain $K_{p,E}$ and integral time constant $T_{i,E}$ generates the $d$-axis current reference, which is limited to $I_{\max}$, and an inner current PI loop ($K_{pd}$, $T_{id}$) tracks this reference. The inverter regulates the load voltage with an AC-voltage PI loop ($K_{p,V}$, $T_{i,V}$), whose reference is lowered through a droop gain $k_{\mathrm{dr}}$ when the DC-link voltage falls. These nine control parameters form the decision vector $\mathbf{x}$ for tuning: 
\begin{align}
\mathbf{x}:=\big[K_{p,\mathrm{PLL}},K_{i,\mathrm{PLL}},K_{p,E},T_{i,E},K_{pd},T_{id},k_{\mathrm{dr}},K_{p,V},T_{i,V}\big]^{\top}\in\mathbb{R}^{9},
\end{align}
while their detailed roles are described in \cite{yazdani2010,tamu_model}.

\subsection{Protection and Low-Voltage Ride-Through Process}\label{sec:prot}

A voltage sag is a short-duration reduction in the grid voltage at the  interconnection point. We denote by $v$ the grid voltage magnitude in per unit (pu) during the voltage sag. As shown in Fig.~\ref{fig:frame}(a), 
the UPS controllers respond to the sag while the protection monitors the rectifier input current and the DC-link voltage. The model includes two protection functions. Overcurrent (OC) protection trips if the  input current $I_{\mathrm{rms}}$ remains above the threshold $I_{\mathrm{OC}}$ for a prescribed delay. DC-link undervoltage (UV) protection trips if the DC-link voltage $V_{dc}$ remains below the threshold $V_{\mathrm{UV}}$ for a prescribed delay. Either trip disconnects the data center load from the grid.

For a sag with retained grid voltage $v$ and the UPS control parameter vector $\mathbf{x}$, the ride-through time $T(v;\mathbf{x})$ is the elapsed time from sag onset to the first OC or UV trip, as shown in Fig.~\ref{fig:frame}(b).
The control parameters $\mathbf{x}$ shape the transient responses of the input current and DC-link voltage, influencing which protection function trips first and when it trips, thus affecting the ride-through time.
Evaluating $T(v;\mathbf{x})$ across the voltage range of interest yields the UPS ride-through curve in Fig.~\ref{fig:frame}(c). The area $A$ to the left of this curve is an aggregate metric that measures the LVRT capability. For a fixed voltage range, a larger $A$ indicates a longer average ride-through time. 



%% file: Problem_Formulation.tex
\section{Problem Formulation and Optimization Algorithm}\label{sec:form}

This section formulates LVRT enhancement as an optimization over the control parameters $\mathbf{x}$ that maximizes the ride-through area $A$, and then develops the BPZO-Adam algorithm to solve it in a model-free manner.


\subsection{Optimization Problem}\label{sec:obj}

For each retained grid voltage $v$, we apply a voltage sag of prescribed duration and record the time $T(v;\mathbf{x})$ from sag onset to the first protection trip. If no trip occurs during the sag, we set $T(v;\mathbf{x})$ to the prescribed sag duration. We approximate the area $A(\mathbf{x})$ to the left of the ride-through curve by a weighted sum over the sampled voltage set $\mathcal{V}$ and maximize it over the admissible control parameters:
\begin{subequations}\label{eq:prob}
\begin{align}
\text{obj.}\ &\max_{\mathbf{x}}\; A(\mathbf{x})=\sum_{v\in\mathcal{V}}\Delta v\, T(v;\mathbf{x}), \label{eq:obj}\\
\text{s.t.}\ \, &\underline{\mathbf{x}}\le\mathbf{x}\le\overline{\mathbf{x}}, \label{eq:box}
\end{align}
\end{subequations}
where $\mathbf{x}\in\mathbb{R}^n$ collects the $n$ control parameters, and $\underline{\mathbf{x}}$ and $\overline{\mathbf{x}}$ are the corresponding lower and upper bounds. 
Here, $\Delta v$ denotes the spacing between consecutive sampled voltage levels. For example, with $\Delta v=0.1$ pu, the sampled voltage levels are $\mathcal{V}\coloneqq\{0,0.1,0.2,\ldots,1.0\}$ pu, spanning the range from zero to nominal grid voltage.


%% file: Algorithm_Design.tex
\subsection{Optimization Algorithm: Batched Projected Zeroth-Order Optimization with Adaptive Moment Estimation}\label{sec:zo}

The problem \eqref{eq:prob} is a black-box optimization as the vendor typically does not disclose the internal details of the UPS model. As a result, we evaluate the ride-through area $A(\mathbf{x})$ using simulations of the vendor-supplied black-box model. 
To solve \eqref{eq:prob}, we develop BPZO-Adam, which extends ZO-Adam by estimating gradients from batches of objective evaluations to reduce variance, projecting each iterate onto the feasible set $\mathcal{X}=\{\mathbf{x}:\underline{\mathbf{x}}\le\mathbf{x}\le\overline{\mathbf{x}}\}$ to ensure feasibility, and the Adam update is applied to accelerate convergence \cite{kingma2015}.


\emph{Batch gradient estimation:} At each iteration $k$, the gradient of $A(\mathbf{x})$ is estimated from a batch of perturbed function evaluations in $N$ random directions \cite{xu_zoadam,liu2020}:
\begin{equation}
\hat{\mathbf{g}}_k=\frac{n}{N}\sum_{j=1}^{N}\frac{A(\mathbf{x}_k+\delta\mathbf{u}_j)-A(\mathbf{x}_k-\delta\mathbf{u}_j)}{2\delta}\,\mathbf{u}_j ,
\label{eq:zo}
\end{equation}
where $\delta>0$ is the smoothing radius, and each random direction $\mathbf{u}_j$ is drawn uniformly from the unit sphere $\mathbb{S}^{n-1}$. Averaging over $N$ directions reduces the estimation variance to obtain a better gradient estimate.

\emph{Projected parameter update:} Since \eqref{eq:prob} is a maximization problem, the parameters move along the estimated gradient (gradient ascent) with the Adam moment estimates:
\begin{subequations}\label{eq:adamall}
\begin{align}
\mathbf{m}_{k+1} &= \beta_1 \mathbf{m}_{k} + (1-\beta_1)\,\hat{\mathbf{g}}_k, \\
\mathbf{s}_{k+1} &= \beta_2 \mathbf{s}_{k} + (1-\beta_2)\,\hat{\mathbf{g}}_k^{2}, \\
\mathbf{x}_{k+1} &= \Pi_{\mathcal{X}}\!\left(\mathbf{x}_k + \alpha\,\frac{\hat{\mathbf{m}}_{k+1}}{\sqrt{\hat{\mathbf{s}}_{k+1}}+\varepsilon}\right),
\end{align}
\end{subequations}
where $\mathbf{m}_{k+1}$ and $\mathbf{s}_{k+1}$ are the first- and second-moment estimates of the gradient. $\alpha$ is the step size, $\beta_1,\beta_2\in[0,1)$ are the exponential decay rates, and $\hat{\mathbf{m}}_{k+1}=\mathbf{m}_{k+1}/(1-\beta_1^{k+1})$ and $\hat{\mathbf{s}}_{k+1}=\mathbf{s}_{k+1}/(1-\beta_2^{k+1})$ are the corresponding bias-corrected estimates, which compensate for the zero initialization $\mathbf{m}_0=\mathbf{s}_0=\mathbf{0}$ \cite{kingma2015}. 
The constant $\varepsilon>0$ prevents division by zero. The projection $\Pi_{\mathcal{X}}$ onto the feasible set $\mathcal{X}=\{\mathbf{x}:\underline{\mathbf{x}}\le\mathbf{x}\le\overline{\mathbf{x}}\}$ clips each updated parameter to its admissible interval, ensuring that each projected iterate satisfies the parameter bounds.


\begin{algorithm}[t]
\caption{BPZO-Adam with Simulation in the Loop for UPS Control Parameter Optimization}
\label{alg:zo}
\begin{algorithmic}[1]
\State \textbf{Input:} grid voltage sag set $\mathcal{V}$, initial parameters $\mathbf{x}_0$, bounds $[\underline{\mathbf{x}},\overline{\mathbf{x}}]$, number of directions $N$, number of iterations $K$, smoothing radius $\delta$, step size $\alpha$, parameters $\beta_1,\beta_2\in[0,1)$ and $\varepsilon>0$.
\State \textbf{Initialize:} moments $\mathbf{m}_0=\mathbf{0}$, $\mathbf{s}_0=\mathbf{0}$; simulate all voltage sags in $\mathcal{V}$ at $\mathbf{x}_0$ to obtain $A(\mathbf{x}_0)$; $\mathbf{x}^\star\gets\mathbf{x}_0$.
\For{$k=0,1,\dots,K-1$}
\State \textbf{(1) Perform Parallel Simulations:}
\For{$j=1,\dots,N$ in parallel}
\State Sample random direction $\mathbf{u}_j\sim\mathrm{Unif}(\mathbb{S}^{n-1})$.
\State \Wrap[2]{Simulate the voltage sag at each $v\in\mathcal{V}$ and $\mathbf{x}_k\pm\delta\mathbf{u}_j$, record $T(v)$, and obtain $A(\mathbf{x}_k\pm\delta\mathbf{u}_j)$ by \eqref{eq:obj}.}
\EndFor
\State \textbf{(2) Gradient Estimation:} compute $\hat{\mathbf{g}}_k$ by \eqref{eq:zo}.
\State \textbf{(3) Optimization Update:}
\State \Wrap[1]{Update $\mathbf{m}_{k+1}$, $\mathbf{s}_{k+1}$ and $\mathbf{x}_{k+1}$ by \eqref{eq:adamall}; simulate all voltage sags at $\mathbf{x}_{k+1}$ to obtain $A(\mathbf{x}_{k+1})$.}
\State \textbf{(4) Best-Point Record:}
\State \Wrap[1]{If $A(\mathbf{x}_{k+1})>A(\mathbf{x}^\star)$, set $\mathbf{x}^\star\gets\mathbf{x}_{k+1}$.}
\EndFor
\State \textbf{Output:} optimized parameters $\mathbf{x}^\star$.
\end{algorithmic}
\end{algorithm}


Algorithm~\ref{alg:zo} summarizes the proposed solution procedure. At each iteration, the $2N$ perturbed points   $\mathbf{x}_k\pm\delta\mathbf{u}_j$ are evaluated at all $|\mathcal{V}|$ voltage sag levels. These $2N|\mathcal{V}|$ simulations are independent and can be conducted fully in parallel.  With sufficient computational resources, the wall-clock time per iteration can approach that of a single
simulation. In addition, convergence bounds for related zeroth-order adaptive momentum methods have been established under specific smoothness and gradient-estimation assumptions \cite{chen2019zoadamm}. These results provide theoretical context for the proposed method, whose convergence behavior on \eqref{eq:prob} is assessed empirically.



%% file: Case_Study_and_Results.tex
\section{Case Study}\label{sec:case}

\subsection{Simulation Setup}\label{sec:sim}

We evaluate the proposed method using the TAMU-ERCOT AI data center EMT model \cite{tamu_model} in PSCAD with a simulation time step of $50~\mu\mathrm{s}$. As shown in Fig.~\ref{fig:ups}, the UPS is connected to a 480-V bus and supplies a 1-MW IT load. Its DC link has a nominal voltage of 0.9~kV and a capacitance of 20~mF, and the rectifier input current limit is $I_{\max}=1.05$~pu. The rectifier disconnects from the grid if either protection function trips. OC protection trips when $I_{\mathrm{rms}}$ remains above 1.1~pu for 10~ms, and UV protection trips when $V_{\mathrm{dc}}$ remains below 0.8~pu for 16.7~ms.
To evaluate LVRT performance, we apply a three-phase voltage sag at the 480-V bus for each retained voltage $v\in\mathcal{V}=\{0,0.1,\ldots,0.9,1.0\}$~pu The sag lasts 2~s for $v\geq 0.8$~pu and 1~s otherwise. We use BPZO-Adam to optimize the $n=9$ control parameters within the bounds in Table~\ref{tab:params}, with $K=50$, $N=2$, $\delta=0.3$, $\alpha=0.2$, $\beta_1=0.5$, and $\beta_2=0.9$. 


\begin{table}[t]
\caption{Control Parameters: Bounds, Initial and Optimized Values}
\label{tab:params}
\centering
\setlength{\tabcolsep}{4.5pt}
\begin{tabular}{llccc}
\toprule
Loop & Parameter & Bounds & Initial & Optimized \\
\midrule
\multirow{2}{*}{Rect. PLL} & $K_{p,\mathrm{PLL}}$ & [200, $10^4$] & 1000 & 1196 \\
 & $K_{i,\mathrm{PLL}}$ & [500, $10^4$] & 3200 & 7620 \\
\multirow{2}{*}{Rect. DC-voltage PI} & $K_{p,E}$ & [1, 100] & 8 & 12.5 \\
 & $T_{i,E}$ (s) & [0.002, 0.05] & 0.01 & 0.016 \\
\multirow{2}{*}{Rect. current PI} & $K_{pd}$ & [0.02, 2] & 0.02 & 0.143 \\
 & $T_{id}$ (s) & [0.02, 0.2] & 0.2 & 0.105 \\
Inv. droop & $k_{\mathrm{dr}}$ & [1.5, 4] & 2 & 3.99 \\
\multirow{2}{*}{Inv. AC-voltage PI} & $K_{p,V}$ & [0.5, 20] & 2 & 3.98 \\
 & $T_{i,V}$ (s) & [0.005, 0.2] & 0.05 & 0.005 \\
\midrule
\multicolumn{2}{l}{Area $A$ (pu$\cdot$s)} & & 0.442 & 0.489 \\
\bottomrule
\end{tabular}
\end{table}

\subsection{LVRT Capability Before and After Optimization}\label{sec:lvrt}
\begin{figure}[t]
\centering
\includegraphics[width=0.85\columnwidth]{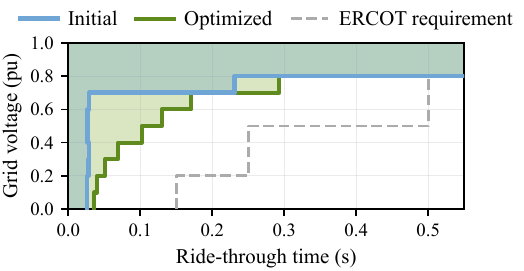}
\caption{Low-voltage ride-through curves with the initial and optimized control parameters, compared with the ERCOT requirement.}
\label{fig:cap}
\end{figure}

Fig.~\ref{fig:cap} compares the LVRT curves obtained with the initial and optimized UPS control parameters. Optimizing the converter control parameters increases the ride-through area $A$ from 0.442 to 0.489~pu$\cdot$s, an improvement of 10.6\%. The gains are concentrated at moderate voltage sags. At retained voltages of 0.6 and 0.5~pu, the ride-through time substantially increases from approximately 30~ms to 170 and 130~ms, respectively. The trip mechanism also changes. With the initial settings, OC protection trips at 0.3--0.7~pu, and UV protection trips at lower retained voltages. With the optimized settings, all observed trips are due to UV protection.


\begin{figure}[t]
\centering
\includegraphics[width=0.8\columnwidth]{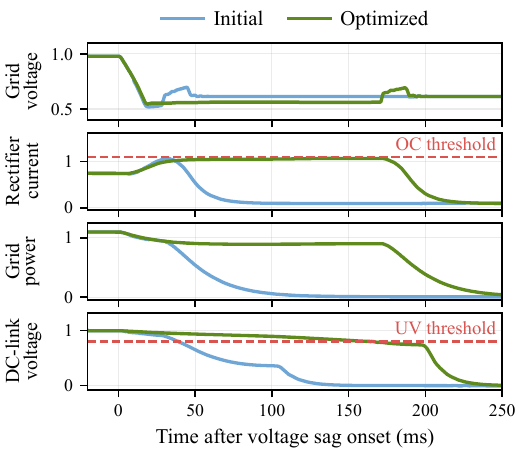}
\caption{Response to a 0.6-pu voltage sag with the initial and optimized control parameters. All quantities are in p.u. Red dashed lines: OC and UV thresholds.} 
\label{fig:wave}
\end{figure}

Fig.~\ref{fig:wave} compares the UPS responses to a 0.6-pu voltage sag. With the initial settings ($K_{pd}=0.02$), the DC-link voltage controller requests more input current as the grid voltage falls. The current overshoots the OC threshold $I_{\mathrm{OC}}$, triggering a trip approximately 30~ms after sag onset. With the optimized settings, the proportional gain of the current loop increases  to $K_{pd}=0.143$. The input current reaches $I_{\max}$ without overshoot, and the UPS continues to draw power near the level permitted by the current limit while the DC link gradually discharges. This comparison shows that controller tuning can avoid the early OC trip under the same current limit. 
The optimization also raises the droop coefficient $k_{\mathrm{dr}}$ to its upper bound. The increased droop lowers the UPS output voltage as the DC-link voltage declines. 

Nevertheless, at retained voltages of 0.7~pu and below, the optimized UPS ride-through times remain shorter than those required by ERCOT. These results illustrate the limits of control tuning alone. Other power support, e.g., from the UPS battery, could supplement the power available from the grid during these sags and extend ride-through~\cite{xie2026enhancing}.

\subsection{Convergence Analysis}\label{sec:conv}
\begin{figure}[t]
\centering
\includegraphics[width=0.8\columnwidth]{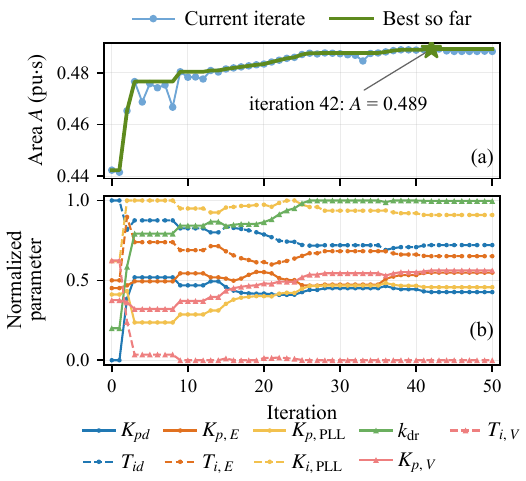}
\caption{Convergence of BPZO-Adam: (a) area $A$; (b) normalized control parameters of the best point (circles: rectifier; triangles: inverter).}
\label{fig:conv}
\end{figure}

Fig.~\ref{fig:conv}(a) shows the convergence of the objective $A(\mathbf{x})$ during the optimization. The ride-through area increases rapidly within the first ten iterations, then improves gradually and levels off after about 40 iterations, reaching its best value of 0.489~pu$\cdot$s at iteration 42. Fig.~\ref{fig:conv}(b) shows the corresponding normalized control parameters, where 0 and 1 denote the lower and upper bounds. The current-loop parameter $K_{pd}$ and the AC-voltage integral parameter $T_{i,V}$ change markedly within the first few iterations, after which $k_{\mathrm{dr}}$ increases steadily to its upper bound, and the remaining parameters settle after about 30 iterations. 
This sequence is consistent with the trip mechanisms described in Section~\ref{sec:lvrt}. The early increase in $K_{pd}$ suppresses input-current overshoot and helps prevent OC trips during moderate voltage sags, contributing to the rapid initial increase in $A$. Once DC-link undervoltage becomes the limiting trip mechanism, further gains come from slowing DC-link discharge. A larger $k_{\mathrm{dr}}$ lowers the UPS output-voltage reference as the DC-link voltage declines, reducing IT load power in the model. A smaller $T_{i,V}$ helps the output voltage track this reference more quickly.


\subsection{Effect of DC-Link Capacitance}\label{sec:cap}
\begin{figure}[t]
\centering
\includegraphics[width=0.8\columnwidth]{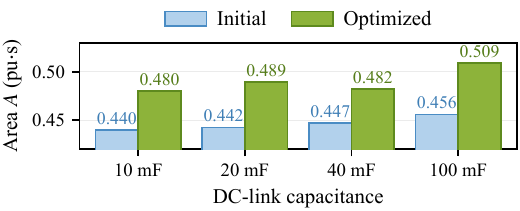}
\caption{LVRT areas $A$ under different DC-link capacitance values with the initial and optimized control parameters.}
\label{fig:capC}
\end{figure}

Increasing the DC-link capacitance allows the capacitor to supply the power deficit during a voltage sag for longer, delaying the UV trip. Fig.~\ref{fig:capC} compares the ride-through area $A$ for capacitances of 10, 20, 40, and 100~mF using the initial and optimized control parameters. We repeat the optimization for each capacitance with $K=50$ iterations.
With the initial settings, increasing the capacitance from 10 to 100~mF raises $A$ only from 0.440 to 0.456~pu$\cdot$s because early OC trips limit the benefit of the additional stored energy. Control optimization increases $A$ by 8--12\% relative to the initial settings at each capacitance, reaching 0.509~pu$\cdot$s at 100~mF. The optimized 10-mF case achieves a larger ride-through area than the 100-mF case with the initial settings (0.480 versus 0.456~pu$\cdot$s). Thus, in this case, tuning the controller yields a larger LVRT gain than increasing the capacitance tenfold.



%% file: Conclusion.tex
\section{Conclusion}\label{sec:concl}

This paper presented BPZO-Adam, a model-free optimization algorithm for improving the LVRT capability of an AI data center UPS by tuning its existing control parameters without hardware upgrades. In EMT simulations of the TAMU-ERCOT AI data center model, the method significantly increased the ride-through area and extended ride-through times at retained voltages of 0.5 and 0.6~pu to approximately four to six times their initial values, primarily by preventing early overcurrent trips. The optimized 10-mF case also achieved a larger ride-through area than the 100-mF case with the initial settings, supporting the use of controller tuning before sizing additional storage. For deeper voltage sags, the optimized settings still fell short of the required ride-through durations. Future work will jointly optimize the UPS control and leverage battery energy storage to address this shortfall.
